\documentclass{styles/svproc}
\usepackage{epsfig}
\usepackage{amsmath}
\usepackage{titlesec}
\usepackage{siunitx}
\usepackage{url}
\usepackage{multirow}

\usepackage[british]{babel}

\graphicspath{{images/}}
\titlespacing*{\section}{0pt}{0.2cm}{0.2cm}

\begin{document}
\mainmatter              
\title{Direct Numerical Simulations of K-type transition in a flat-plate boundary layer with supercritical fluids}
\titlerunning{K-type transition in a flat-plate boundary layer with supercritical fluids}  

\author{Pietro Carlo Boldini\inst{1} \and Benjamin Bugeat\inst{2} \and Jurriaan W.R.~Peeters \inst{1} \and \\Markus Kloker \inst{3} \and Rene Pecnik \inst{1}}
\authorrunning{P.C.~Boldini et al.} 
\tocauthor{P.~C.~Boldini, B.~Bugeat, J.~W.~R.~Peeters, M.~Kloker, R.~Pecnik}
\institute{Process and Energy Department, Delft University of Technology, Leeghwaterstraat 39, 2628 CB Delft, The Netherlands,\\
\email{pietro.c.boldini@gmail.com, r.pecnik@tudelft.nl}\\
\and
School of Engineering, University of Leicester, University Road, Leicester, LE1 7RH, United Kingdom,\\
\and
Institute of Aerodynamics and Gas Dynamics, University of Stuttgart, Pfaffenwaldring 21, 70569 Stuttgart, Germany}

\maketitle              

\vspace{-0.6cm}
\begin{abstract}
We investigate the controlled K-type breakdown of a flat-plate boundary-layer with highly non-ideal supercritical fluid at a reduced pressure of $p_{r,\infty}=1.10$. Direct numerical simulations are performed at a Mach number of $M_\infty=0.2$ for one subcritical (liquid-like regime) temperature profile and one strongly-stratified transcritical (pseudo-boiling) temperature profile with slightly heated wall. In the subcritical case, the formation of aligned $\Lambda$-vortices is delayed compared to the reference ideal-gas case of Sayadi et al.~(\textit{J.~Fluid Mech.}, vol.~724, 2013, pp.~480–509), with steady longitudinal modes dominating the late-transitional stage. When the wall temperature exceeds the pseudo-boiling temperature, streak secondary instabilities lead to the simultaneous development of additional hairpin vortices and near-wall streaky structures near the legs of the primary aligned $\Lambda$-vortices. Nonetheless, transition to turbulence is not violent and is significantly delayed compared to the subcritical regime.
\end{abstract}
\vspace{-0.4cm}
\section{Introduction}

When a fluid at supercritical pressure is isobarically heated or cooled across the pseudo-boiling (Widom) line, it undergoes a continuous phase transition characterised by strong thermophysical property variations, which can impact the transition to turbulence. Research on the hydrodynamic stability of boundary layers with fluids exhibiting non-ideal thermodynamic behaviour has recently gained interest  \cite{Robinet1}. The newly identified mode-II instability \cite{Ren1} in the transcritical regime, associated with the minimum of kinematic viscosity at the Widom line, arises from the interaction between shear and baroclinic waves \cite{Bugeat1}. Non-modal stability analysis revealed that mode II alters the streamwise invariance of optimal streamwise streaks \cite{Boldini1}. In three-dimensional boundary-layer flows, the inviscid mode-II instability prevailed over cross-flow instability despite flow acceleration \cite{Ren2}. 
While these studies focused on the linear regime, non-linear interactions and breakdown to turbulence remain unexplored. As a first attempt, Boldini et al.~\cite{Boldini2} investigated the H-type breakdown of a flat-plate boundary layer under subcritical and transcritical temperature conditions. They showed that staggered $\Lambda$-vortices with inverted $\Lambda$-shape at their valleys (secondary vortex systems) lead to a violent breakdown process. Here, we examine the other main controlled transitional scenario, the K-type breakdown \cite{Rist1}, by means of direct numerical simulation (DNS).

\section{Methodology}
The non-ideal compressible, high-order finite-difference code CUBENS is employed for the simulations in this work; see Ref.~\cite{Boldini3} for details. 
A rectangular computational domain is selected containing the boundary layer of a flat-plate, (see Ref.~\cite{Boldini2}), with the inlet location at $x_0$ from the leading edge chosen based on the inlet Reynolds number of $Re_{x,0}=10^5$. The reference length scale is the boundary-layer thickness at the domain inlet, $\delta^*_{99}(x_0)=\delta^*_{99,0}$. Non-reflecting boundary conditions with sponge zones are applied at the inlet, top, and outflow boundaries, while no-slip and fully-reflective conditions are enforced at the wall. The grid is equidistant in the streamwise and periodic spanwise directions, with grid stretching applied in the wall-normal direction.

Mimicking the ideal-gas case of Ref.~\cite{Sayadi1}, disturbances are introduced at the wall via a blowing/suction disturbance strip with a length of $x/\delta_{99,0}=13$, at $Re_\delta=\sqrt{Re_x}=381$. Here, $\delta$ is the local Blasius length scale defined as $\delta^* = (\mu^*_\infty x^*/\rho^*_\infty/u^*_\infty)^{1/2}$. The wall-normal velocity distribution is specified as 
\begin{equation}
    v(y=0)=f(x) \left[ A_{\text{2-D}} \sin (\omega_0 t) + A_{\text{3-D}} \sin(\omega_{0} t) \cos( \beta_0 z) \right],
    \label{eq:diststrip}
\end{equation}
where $f(x)$ is provided in \cite{Sayadi1}. The amplitudes $A_{\text{2-D}}=2.8 \times 10^{-2}$ and $A_{\text{3-D}}=4.0 \times 10^{-4}$, corresponding to the fundamental two-dimensional (2-D) and the oblique three-dimensional (3-D) wave pair, respectively, share the same frequency, $F_0=\omega_0/Re_{\delta_{99,0}} = 110 \times 10^{-6}$, as in Ref.~\cite{Sayadi1}. The spanwise wavenumber is $\beta_0=2\pi/z_e$. In the following, modes are denoted in a double-spectral notation ($h \omega_0$, $k\beta_0$). 

\section{Cases, base flow and linear stability analysis}
\begin{figure*}[!t]
\centering
\includegraphics[angle=-0,trim=0 0 0 0, clip,width=1.0\textwidth]{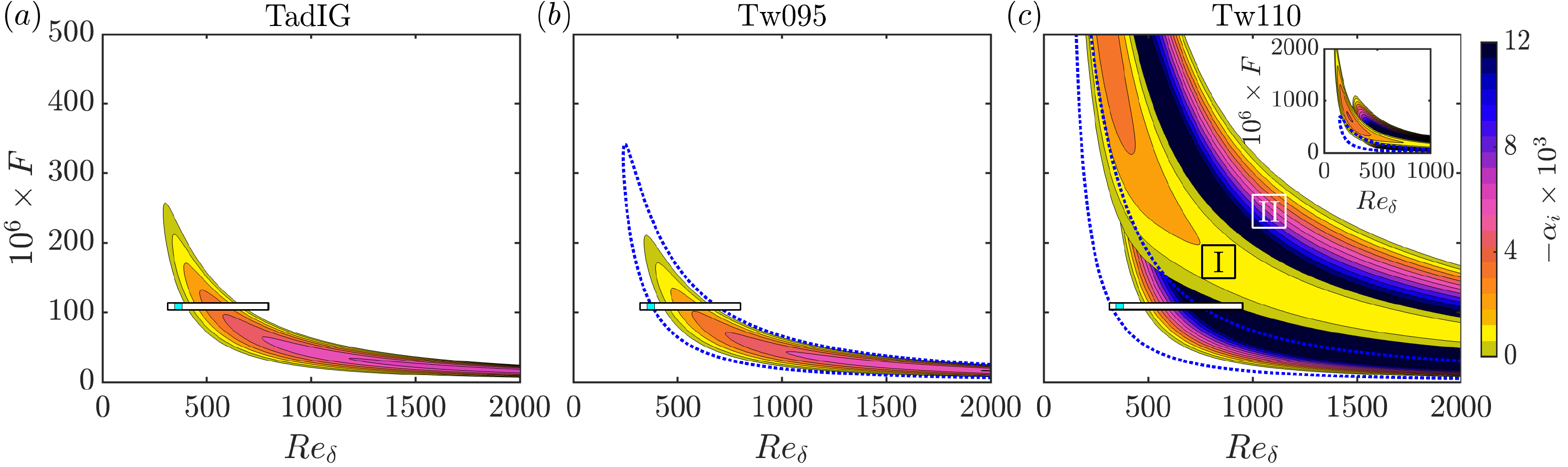} 
\vspace{-0.7cm}
\caption{\label{fig:LST}Growth-rate ($-\alpha_i$) contours in the $Re_\delta$--$F$ stability diagram for 2-D disturbances: (a) TadIG, (b) Tw095, and (c) Tw110 (mode I and II). The dotted blue lines in (b,c) represent the ideal-gas neutral stability with equal $T^*_w/T^*_\infty$-ratio. The inset of (c) highlights the wide frequency band of mode I and II. The DNS domain and perturbation strip for the fundamental breakdown, i.e.~$F_{0}=110\times10^{-6}$, are indicated by the white and cyan bars, respectively. $Re_\delta$ is scaled with the local Blasius length.  }
\end{figure*}

The flow parameters in this study match those used for the DNS investigation of the H-type breakdown in Ref.~\cite{Boldini2}. The supercritical fluid is described by the reduced Van der Waals equation of state, where reduced properties are denoted as $(\cdot)_r=(\cdot)^*/(\cdot)^*_c$, with $(\cdot)^*_c$ representing the critical point values. Transport properties are modelled using the Jossi, Stiel, and Thodos analytical relations \cite{Boldini3}. The free-stream parameters are consistent among all cases: a supercritical pressure of $p_{r,\infty}=1.10$, temperature of $T_{r,\infty}=0.90$, density of $\rho_{r,\infty}=1.805$ (in the liquid-like region), and Mach number of $M_\infty=0.2$. Two reduced wall temperatures are considered: one in the liquid-like regime at $T_{r,w}=0.95$, or $T^*_{w}/T^*_{\infty}=1.056$ (subcritical case, referred to as Tw095), and the other in the vapour-like regime at $T_{r,w}=1.10$, or $T^*_{w}/T^*_{\infty}=1.222$ (transcritical case, referred to as Tw110). For further details on the flow parameters; see Ref.~\cite{Boldini2}, while computational parameters for all cases are presented in Tab.~\ref{tab:DNS}.
\begin{table}[!t]
\caption{\label{tab:DNS}Numerical parameters of the K-type simulations. Viscous units $(\cdot)^+$ are calculated at $\max\{C_f\}$. Case TadIG refers to the ideal-gas reference case of Ref.~\cite{Sayadi1}.}
\begin{center}
\begin{tabular}{c@{\hspace{0.5cm}} c@{\hspace{0.2cm}} c@{\hspace{0.2cm}} c@{\hspace{0.2cm}} c@{\hspace{0.2cm}} c@{\hspace{0.2cm}} c@{\hspace{0.2cm}} c@{\hspace{0.2cm}} c@{\hspace{0.2cm}} c@{\hspace{0.2cm}} c@{\hspace{0.2cm}}}
\hline\rule{0pt}{12pt}
Case & $L_x/\delta_0$ & $L_y/\delta_0$ & $L_z/\delta_0$ & $N_x$ & $N_y$ & $N_z$ & $\Delta x^+$ & $\Delta y^+_{min}$ & $\Delta z^+$ & $Re_{\theta,max}$ \\ 
\hline\rule{0pt}{12pt}
  TadIG & 347 & 20 & 9.63 & 3000 & 600 & 150 & 9.3 & 0.61 & 5.2 & 526 \\
  Tw095 & 352 & 20 & 9.63 & 3000 & 600 & 150 & 10.0 & 0.65 & 5.5 & 513  \\
  Tw110 & 570 & 40 & 9.63 & 8400 & 600 & 150 & 6.1 & 0.68 & 5.8 & 489  \\
\hline\rule{0pt}{12pt}
\end{tabular}
\end{center}
\end{table} 
The ideal-gas K-type breakdown in Ref.~\cite{Sayadi1}, labelled as TadIG, is included for validation and comparison. The laminar boundary-layer profiles are discussed in Ref.~\cite{Boldini2}, with the strongest density and thermophysical property variations occurring near the Widom line for case Tw110. Linear stability theory is used to analyse the linear disturbance behaviour across the entire relevant disturbance frequency range, and the stability diagrams are compared in Fig.~\ref{fig:LST}. The spatial growth rate $-\alpha_i$ is plotted as a function of the local Reynolds number $Re_\delta=\sqrt{Re_{x,0}}/\delta_{99,0}=Re_{\delta,0}/\delta_{99,0}$ and frequency parameter $F$. 
 
Within the considered DNS domain for the K-type breakdown at $F_{0}=110\times10^{-6}$ in Fig.~\ref{fig:LST}(c), the unstable region of mode (1,0) is caused exclusively by mode II. Moreover, the broad frequency bands of instability  for mode I and II are visible in the inset of Fig.~\ref{fig:LST}(c).

\section{Direct numerical simulations}
\begin{figure*}[!t]
\centering
\includegraphics[angle=-0,trim=0 0 0 0, clip,width=1.0\textwidth]{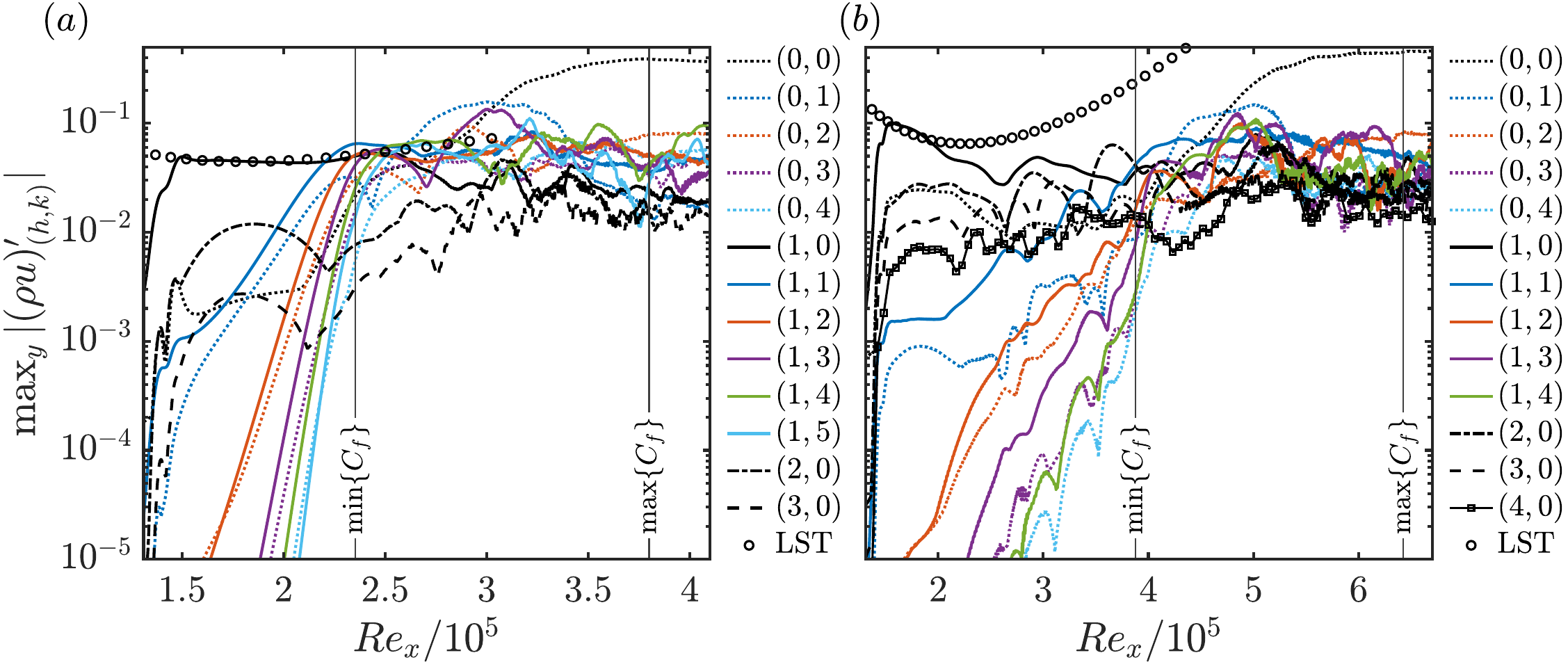} 
\vspace{-0.7cm}
\caption{\label{fig_fft}Streamwise development of the $y$-maximum $(\rho u)^\prime$ disturbance amplitudes: (a) Tw095, (b) Tw110. LST results are shown in black circles. Note that the same $v$-distribution has been applied to the disturbance strip in both cases.}
\end{figure*}

Figure~\ref{fig_fft} shows the downstream development of the maximum amplitudes (over $y$) of $(\rho u)^\prime$ for cases Tw095 (a) and Tw110 (b). In the subcritical regime (Fig.~\ref{fig_fft}(a)), the modal development resembles case TadIG (not shown here). However, due to the increase in wall temperature towards the Widom line, the growth of secondary oblique disturbance waves $(1, \pm 1)$ is weakened, delaying the location of $\min\{C_f\}$. Later, all modes saturate, and from $Re_x/10^5 \approx 2.6$ onward, the non-linearly generated longitudinal mode $(0, 1)$ reaches the highest amplitudes, surpassing the other modes over a considerable streamwise distance. In this case, the higher harmonic mode $(1, 3)$ is destabilised, whereas in case TadIG (not shown here), it is the $(1, 4)$ mode. Further downstream, in the strongly non-linear region, the generated mean-flow distortion $(0, 0)$ reaches its largest amplitude, indicating a strong mean flow deformation due to transition to turbulence, i.e.~$\max\{C_f\}$. In contrast, in the transcritical regime (Fig.~\ref{fig_fft}(b)), the non-linear, high-amplitude primary wave initially saturates due to non-linear effects and is surpassed by its first higher harmonic $(2,0)$ at $Re_x/10^5 \approx 3.4$, where $\Lambda$-structures reveal at halved streamwise wavelength. Note that the higher harmonics ($(2,0)$, $(3,0)$, etc.) are highly unstable according to LST, as shown in Fig.~\ref{fig:LST}(c), are more strongly excited at the disturbance strip than in the subcritical case, and reach a higher amplitude level than $(1, 1)$ and $(0, 1)$ in the early transitional stage. Once $(2,0)$ dominates, the secondary oblique disturbance waves experience rapid streamwise growth around the $\min\{C_f\}$ location (subharmonic resonance), together with the higher harmonic modes ($(1,2)$, $(0,2)$ $(1, 3)$, etc.). In the transitional region, the steady longitudinal mode $(0, 1)$ reaches its highest amplitude from $Re_x/10^5 \approx 4.3$ onwards, prior to the full transition to turbulence. 

\begin{figure*}[!t]
\centering
\includegraphics[angle=-0,trim=0 0 0 0, clip,width=1.0\textwidth]{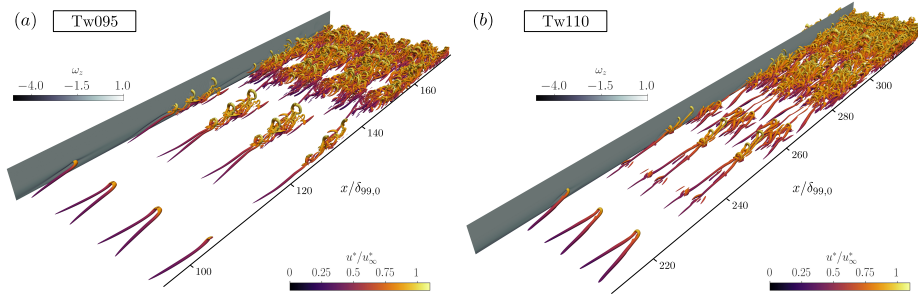} 
\vspace{-0.7cm}
\caption{\label{fig_qvort}Instantaneous isosurfaces of the Q-criterion, coloured by the streamwise velocity $u$: (a) Tw095 ($Q = 0.06$), (b) Tw110 ($Q = 0.016$). The side plane indicates the instantaneous spanwise vorticity $\omega_z$. Note the different x-ranges and that the domain is copied three times in the spanwise direction for better visualisation. }
\end{figure*}

Figure~\ref{fig_qvort} displays the instantaneous flow structures through isocontours of the Q-criterion. In the subcritical regime (Fig.~\ref{fig_qvort}(a)) the sequence of aligned $\Lambda$-structures, with a high-shear layer at their tips (see spanwise vorticity contours), and hairpin-shaped vortices can be clearly observed. As they propagate downstream, $\Lambda$-structures are known to elongate further \cite{Rist1}. This behaviour is even more pronounced here, as $(0, 1)$ is the dominant mode in the region around $x/\delta_{99,0}=120$. In the transcritical regime (Fig.~\ref{fig_qvort}(b)), the breakdown is significantly delayed compared to the subcritical regime (Fig.~\ref{fig_qvort}(a)). In particular, $\Lambda$-structures are aligned, and no secondary vortex structures are found at their valleys, as previously discovered in the H-type breakdown under the same flow conditions \cite{Boldini2}. Moreover, we observe a less rapid and abrupt breakdown to turbulence than in the H-type breakdown. 
Interestingly, in Fig.~\ref{fig_qvort}(b), around $x/\delta_{99,0}=263$ ($Re_x/10^5 \approx 4.7$), a pair of secondary hairpin vortices are observed lifting off the sides of the main hairpin vortex, accompanied by strong vortical near-wall structures. These secondary structures originate farther upstream from the main legs of the $\Lambda$-vortex at $x/\delta_{99,0}=238$ ($Re_x/10^5 \approx 4.35$). In this region, the streamwise-elongated vortex structure is predominantly composed of the $(0,1)$ mode, which has exceeded an amplitude of $10\%$. 
\begin{figure*}[!b]
\centering
\includegraphics[angle=-0,trim=0 0 0 0, clip,width=1.0\textwidth]{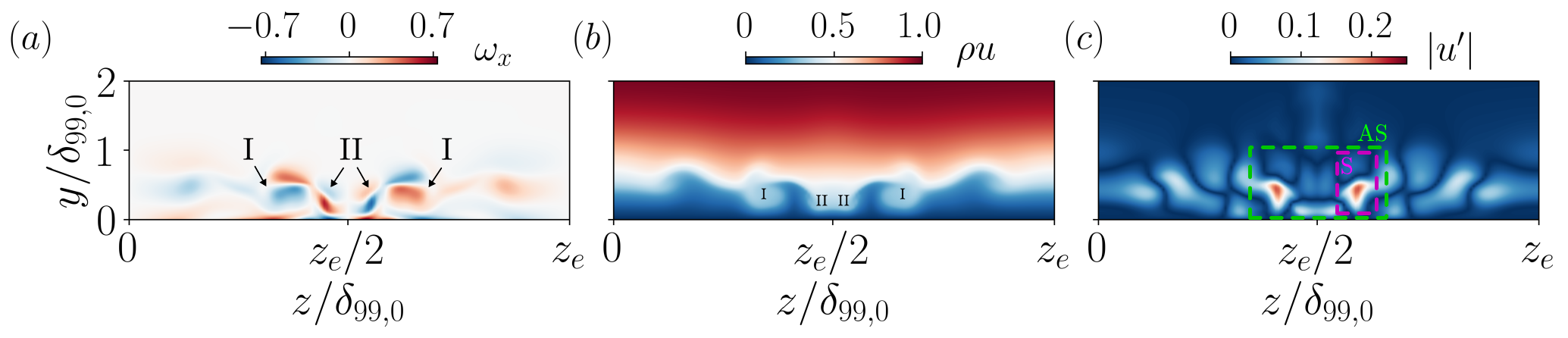} 
\vspace{-0.7cm}
\caption{\label{fig_zcut}Instantaneous isosurfaces at $x/\delta_{99,0}=238$ ($Re_x/10^5 \approx 4.35$) in the $y$-$z$ plane:~(a) streamwise vorticity $\omega_x$ of counter-rotating (I) and co-rotating (II) vortex pairs, (b) streamwise momentum $\rho u$, and (c) absolute value of streamwise velocity perturbation $u'$ with symmetric (S) and antisymmetric (AS) mode.}
\end{figure*}
To highlight its structure, Fig.~\ref{fig_zcut}(a) displays the two main legs of oppositely signed streamwise vorticity $\omega_x$ (symbol I:~one at $z/\delta_{99,0}>z_e/2$ and the other at $z/\delta_{99,0}<z_e/2$). Additionally, two co-rotating legs of $\pm \omega_x$ (symbol II) are noticeable closer to the wall between the main vortex-pair structure, having a higher streamwise momentum $\rho u$ than the surrounding low-velocity, low-density fluid; see Fig.~\ref{fig_zcut}(b). The interaction between counter-rotating and co-rotating vortex tubes generates significant spanwise modulation, with low-speed regions, i.e.~low-speed streaks, forming between the vortices of opposite vorticity signs at $z/\delta_{99,0}>z_e/2$ and $z/\delta_{99,0}<z_e/2$. Given the aforementioned non-linear amplitude, these streaks undergo a localised secondary instability (LSI), exhibiting two different instability modes on either side of the streak or streak pair. In Fig.~\ref{fig_zcut}(c), the contours of streamwise velocity perturbations over the spanwise distance are shown. At the centre of each low-speed streak, a symmetric varicose mode is detected, which develops into the aforementioned symmetric hairpin vortex. Simultaneously, an antisymmetric sinuous mode is identified, causing the streak pair to meander with respect to $z/\delta_{99,0}=z_e/2$, and generating additional near-wall streamwise vortices downstream. 

\section{Conclusions}
The K-type breakdown of a weakly wall-heated flat-plate boundary layer at a Mach number of $0.2$ with a fluid at supercritical pressure has been investigated. In the subcritical heating regime, where $T^*<T^*_{pc}$, breakdown to turbulence is slightly delayed compared to the reference ideal-gas case in Ref.~\cite{Sayadi1}. Conversely, in the transcritical heating regime, where $T^*_w>T^*_{pc}$, large-amplitude higher harmonics dominate the initial breakdown stage, delaying both the location and strength of the fundamental resonance. Furthermore, subharmonic resonance between mode $(2,0)$ and $(1,1)$ is active just before the transition onset. Farther downstream, a competition between symmetric and antisymmetric modes of localised secondary instability emerges near the legs of the primary aligned $\Lambda$-vortices at the peak stations. Unlike the rapid and violent H-type breakdown in the transcritical wall-heating regime, the transcritical K-type breakdown is more gradual and does not exhibit strong secondary vortical structures at the spanwise valley positions.

%
%

\end{document}